 \documentclass[final,authoryear,5p,times,twocolumn]{elsarticle}

\usepackage{amssymb}

\journal{Icarus}
\bibpunct{(}{)}{;}{a}{}{,}
\usepackage{graphicx,subfigure,longtable}

\usepackage{endfloat}

\usepackage{ulem}
\usepackage{hyperref}

\newcommand{\degr}{\ensuremath{^\circ}}

\begin{document}

\begin{frontmatter}

\title{Unexpected D-type Interlopers in the Inner Main Belt}

\author[har,mit,hub]{Francesca E. DeMeo}
\author[mit]{Richard P. Binzel}
\author[imcce,esac]{Beno{\^i}t Carry}  
\author[mit]{David Polishook}
\author[mit]{Nicholas A. Moskovitz}

\address[har]{Harvard-Smithsonian Center for Astrophysics, 60 Garden Street, MS-16, Cambridge, MA, 02138, USA. fdemeo@cfa.harvard.edu}
\address[mit]{Department of Earth, Atmospheric, and Planetary Sciences, Massachusetts Institute of Technology, 77 Massachusetts Avenue, Cambridge, MA 02139 USA}
\address[hub]{Hubble Fellow}
\address[imcce]{Institut de M{\'e}canique C{\'e}leste et de Calcul des {\'E}ph{\'e}m{\'e}rides, Observatoire de Paris, UMR8028 CNRS, 77 av. Denfert-Rochereau 75014 Paris, France}
\address[esac]{European Space Astronomy Centre, ESA, P.O. Box 78, 28691 Villanueva de la Ca{\~n}ada, Madrid, Spain}

\begin{abstract}
  Very red featureless asteroids (spectroscopic D-types) are expected to have formed in the outer 
  solar system far from the sun. They comprise the majority of asteroids in
  the Jupiter Trojan population, and are also commonly found in the outer main belt and among Hildas. 
  The first evidence for D-types in the inner and middle parts of the main belt was seen in the Sloan Digital
  Sky Survey (SDSS).  Here we report follow-up observations of SDSS D-type candidates in
  the near-infrared. Based on follow up observations of 13 SDSS D-type candidates, we find a $\sim$20\% positive
  confirmation rate. Known inner belt D-types range in diameter from roughly 7 to 30 kilometers. 
  Based on these detections we estimate there are $\sim$100 inner belt D-types with diameters between 2.5 and 20km.
  The lower and upper limits for total mass of inner belt D-types is 2x10$^{16}$kg to 2x10$^{17}$kg which represents 0.01\% to 0.1\% 
  of the mass of the inner belt.
  The inner belt D-types have albedos at or above the upper end typical for D-types which raises the question
  as to whether these inner belt bodies represent only a subset of D-types, they have been altered by external factors such 
  as weathering processes, or if they are compositionally distinct from other D-types. All D-types and candidates 
  have diameters less than 30km, yet there is no obvious parent body in the inner belt.
  Dynamical models have yet to show how D-types originating from the outer solar system could penetrate into the inner reaches 
  of the Main Belt under current scenarios of planet formation and subsequent Yarkovsky drift.
\end{abstract}

\begin{keyword}
ASTEROIDS \sep SPECTROSCOPY 

\end{keyword}

\end{frontmatter}


\section{Introduction}

  Spectral D-type asteroids are defined by their very red spectral slopes and
  typical lack of distinguishing absorption features in the visible and
  near-infrared wavelength ranges
  \citep{Tholen1984,Bus2002b,DeMeo2009taxo}. Despite a lack of clear absorption
  bands to determine their composition, D-types are expected to be rich in
  organic compounds \citep{Gradie1980, Cruikshank1992}. Spectral 
  surveys of D-types include
  \citet{Lagerkvist1993},
  \citet{Fitzsimmons1994},
  \citet{Fornasier2004},
  \citet{Dotto2006},
  \citet{Fornasier2007},
  \citet{Roig2008b},  
  \citet{Emery2011}, and 
  \citet{Yang2011}. 
  
  Traditionally, D-types have only been found in the outer main belt, beyond 2.82 AU into the Jovian Trojan population at 5.2 AU.
   They are the dominant spectral type among Trojan asteroids and make
  up an important fraction of Hildas, Cybeles, and outer main-belt asteroids
  \citep{Gradie1982,Grav2012,DeMeo2013}. Their spectral slopes are significantly higher than
  for C, and X-types found in the main asteroid belt
  \citep{Tholen1984,Bus2002a,DeMeo2009taxo}, but lower (at visible
  wavelengths) than many of the ultra-red Centaurs and TNOs such as
  (83982) Crantor,
  (42301) 2001 UR$_{163}$, and the prototypical
  (5145) Pholus
  \citep{Alvarez2008,Fornasier2009,DeMeo2009tno,Perna2010,Cruikshank1998}.  

  A few D-types have been found among the NEO population
  \citep{Binzel2004,DeMeo2008}, but they have been conspicuously absent from the
  inner and middle asteroid belt regions ranging from 2.0 to 2.82\,AU
  \citep{Bus1999,Bus2002a}. This absence has fallen in line with dynamical
  models that suggest D-type asteroids formed farther out in the solar system
  and were transported inward by planetary migration but made it only as far as
  the outer main belt \citep{Levison2009, Morbidelli2005}. The inner, middle,
  and outer sections of the main belt, ranging from
  2.0--2.5, 2.5--2.82, and 2.82--3.2\,AU respectively, 
  are separated by the 3:1 and 5:2 mean motion resonances. 

  Evidence for a small number of D-type asteroids in the inner and middle asteroid
  belt has been seen in the Sloan Digital Sky Survey
  \citep[SDSS,][]{Ivezic2001,Ivezic2002} by \citet{Carvano2010} and \citet{DeMeo2013}. In
  this work we follow-up these ``SDSS D-type candidates'' with near-infrared
  spectral measurements and confirm the presence of D-types in the inner
  belt. We present the spectra, determine the positive detection rate of D-type
  asteroids based on SDSS candidates, calculate the total expected D-types in
  the inner belt according to their size distribution and mass, and discuss
  potential dynamical implications.

\section{Observations}

  \subsection{Target Selection}
    Candidate D-types were chosen among objects observed in the Sloan Digital
    Sky Survey (SDSS) Moving Object Catalog (MOC). We use the fourth release
    (MOC4), including observations prior to March 2007. We restrict the sample
    based on data quality. Detail of the sample selection and classification
    can be found in \citet{DeMeo2013}. From this subset we create a list of
    all objects in the inner belt that have been classified at least once as a
    D-type. We then shorten this list further by removing any objects that
    have more than one observation that differs significantly from a D-type,
    such as an object that is classified once as an S and once as a D. Our
    final list includes 79 inner belt SDSS candidate D-types, 11 of which have been
    observed multiple times (Table~\ref{tab:cand}). We also list the three Hungaria and three Phocaea candidates.
    We observed 13 targets using SpeX at the IRTF and FIRE at Magellan. 

\begin{table*}[t]
\caption{Inner Belt D-type Candidates from SDSS MOC4}
\label{tab:cand}
\begin{center}
\begin{tabular}{rlcccr}
  \hline
  \hline
  \multicolumn{2}{c}{Designation} & H & a &  e &  i   \\
  (\#)\footnote{Objects in bold were observed in this work. Objects marked with a * are confirmed D-types.} & (Name) & (mag) & (au) & & (\degr) \\
  \hline
 \multicolumn{3}{l}{Inner belt candidates}   & & \\

  2806 & Graz	&	13.1	&	2.38	&	0.05	&	2.3	\\
  \textbf{3283*} & Skorina	&	12.6	&	2.40	&	0.10	&	6.9	\\
  \textbf{5202} & 1983 XX	&	13.0	&	2.40	&	0.17	&	12.6	\\
  5302 & Romanoserr	&	13.9	&	2.33	&	0.04	&	2.1	\\
  8069 & Benweiss	&	14.0	&	2.34	&	0.07	&	6.1	\\
  8856 & Celastrus	&	14.3	&	2.35	&	0.10	&	2.4	\\
 10573 & Piani	&	14.5	&	2.45	&	0.16	&	14.8	\\
 \textbf{14291} & 1104 T-1	&	13.9	&	2.35	&	0.13	&	11.7	\\
 \textbf{15112*} & Arlenewolf	&	14.2	&	2.30	&	0.15	&	3.8	\\
 \textbf{16025} & 1999 CA104	&	14.1	&	2.29	&	0.21	&	6.8	\\
 \textbf{17906*} & 1999 FG32	&	13.5	&	2.47	&	0.17	&	10.7	\\
 20180 & Annakoleny	&	14.5	&	2.48	&	0.12	&	12.9	\\
 20243 & 1998 DB36	&	13.9	&	2.34	&	0.22	&	11.5	\\
 \textbf{20439} & 1999 JM28	&	13.9	&	2.40	&	0.19	&	1.4	\\
 \textbf{20452} & 1999 KG4	&	14.0	&	2.35	&	0.24	&	7.9	\\
 20749 & 2000 AD199	&	13.7	&	2.45	&	0.15	&	14.3	\\
 21606 & 1999 FH6	&	15.0	&	2.18	&	0.07	&	0.9	\\
 22788 & von Steube	&	13.7	&	2.45	&	0.21	&	8.5	\\
 25718 & 2000 AH170	&	14.7	&	2.29	&	0.11	&	10.9	\\
 27435 & 2000 FZ35	&	14.4	&	2.47	&	0.12	&	12.8	\\
 27842 & 1994 QJ	&	14.2	&	2.35	&	0.23	&	24.9	\\
 29452 & 1997 RV2	&	14.4	&	2.39	&	0.13	&	12.3	\\
 \textbf{33917} & 2000 LK19	&	14.9	&	2.33	&	0.11	&	7.3	\\
 33964 & 2000 NS10	&	14.9	&	2.36	&	0.14	&	6.8	\\
 34692 & 2001 KE61	&	15.0	&	2.31	&	0.25	&	7.2	\\
 35058 & 1985 RP4	&	15.0	&	2.42	&	0.25	&	3.8	\\
 38699 & 2000 QX63	&	15.7	&	2.27	&	0.09	&	7.0	\\
 39809 & Fukuchan	&	16.2	&	2.23	&	0.22	&	5.4	\\
 39881 & 1998 EK11	&	14.7	&	2.36	&	0.19	&	3.9	\\
 44169 & 1998 KK2	&	14.0	&	2.46	&	0.28	&	14.7	\\
 47320 & 1999 XA15	&	13.7	&	2.41	&	0.09	&	22.4	\\
 48049 & 2001 DB90	&	15.1	&	2.47	&	0.13	&	11.0	\\
 48763 & 1997 JZ	&	15.0	&	2.39	&	0.13	&	13.6	\\
 49092 & 1998 RK71	&	15.1	&	2.23	&	0.06	&	6.1	\\
 52570 & 1997 JC1	&	14.7	&	2.38	&	0.13	&	13.8	\\
 55391 & 2001 ST277	&	14.6	&	2.39	&	0.20	&	11.8	\\
 55567 & 2002 CS6	&	13.5	&	2.32	&	0.27	&	22.9	\\
 55590 & 2002 PB97	&	15.3	&	2.34	&	0.25	&	3.9	\\
 57546 & 2001 TO21	&	15.7	&	2.44	&	0.09	&	6.6	\\
 58005 & 2002 TR207	&	15.5	&	2.37	&	0.07	&	10.3	\\
 58684 & 1998 AA11	&	15.8	&	2.42	&	0.16	&	3.5	\\
 65915 & 1998 FO34	&	14.2	&	2.34	&	0.09	&	12.9	\\
 68004 & 2000 XD38	&	15.2	&	2.44	&	0.11	&	14.5	\\
 68448 & Sidneywolf	&	15.3	&	2.29	&	0.24	&	9.2	\\
 73598 & 2912 T-2	&	17.0	&	2.23	&	0.14	&	1.5	\\
 76973 & 2001 BT53	&	15.0	&	2.36	&	0.15	&	11.9	\\
 80532 & 2000 AV71	&	15.9	&	2.29	&	0.04	&	4.5	\\
 84480 & 2002 TM266	&	14.8	&	2.30	&	0.26	&	11.1	\\
 84802 & 2002 YC1	&	14.8	&	2.36	&	0.22	&	9.9	\\
 85054 & 6841 P-L	&	16.4	&	2.31	&	0.12	&	4.0	\\
 96199 & 1992 EY24	&	15.8	&	2.20	&	0.12	&	4.0	\\
 97965 & 2000 QW143	&	15.7	&	2.36	&	0.14	&	14.3	\\
 99449 & 2002 CJ30	&	16.4	&	2.29	&	0.11	&	4.8	\\
104163 & 2000 EL76	&	15.9	&	2.36	&	0.11	&	12.3	\\
109253 & 2001 QT103	&	16.0	&	2.48	&	0.22	&	14.9	\\
111899 & 2002 FD11	&	16.5	&	2.21	&	0.07	&	4.0	\\
122596 & 2000 RG35	&	14.5	&	2.43	&	0.11	&	14.7	\\
123240 & 2000 UU59	&	15.8	&	2.32	&	0.25	&	5.9	\\
\textbf{125102} & 2001 UH35	&	15.3	&	2.31	&	0.08	&	11.1	\\
125353 & 2001 VA61	&	15.6	&	2.39	&	0.18	&	12.4	\\
125804 & 2001 XQ158	&	16.6	&	2.38	&	0.04	&	2.4	\\
129761 & 1999 GL6	&	14.9	&	2.30	&	0.24	&	10.8	\\
135254 & 2001 SY43	&	16.2	&	2.35	&	0.19	&	4.8	\\
139039 & 2001 EP3	&	16.2	&	2.28	&	0.13	&	10.7	\\
155597 & 2000 CE61	&	16.4	&	2.28	&	0.08	&	7.7	\\
172458 & 2003 RF4	&	16.5	&	2.32	&	0.13	&	5.5	\\
172548 & 2003 UA80	&	16.5	&	2.45	&	0.13	&	6.1	\\
194949 & 2002 AW160	&	16.2	&	2.37	&	0.15	&	7.2	\\
207564 & 2006 PZ	&	16.1	&	2.45	&	0.18	&	11.2	\\
209988 & 2006 HL86	&	17.1	&	2.26	&	0.09	&	5.6	\\
210531 & 1999 FB1	&	16.5	&	2.22	&	0.15	&	5.9	\\
215706 & 2004 AP10	&	15.4	&	2.46	&	0.28	&	15.0	\\
218552 & 2005 EM28	&	16.1	&	2.31	&	0.12	&	10.2	\\
218612 & 2005 PY3	&	15.4	&	2.41	&	0.23	&	15.7	\\
\textbf{224306} & 2005 UV8	&	16.1	&	2.36	&	0.23	&	6.8	\\
231103 & 2005 SM70	&	17.4	&	2.32	&	0.24	&	2.4	\\
239887 & 2000 QU163	&	17.0	&	2.3	&	0.15	&	12.6	\\
\textbf{247264} & 2001 SW8	&	16.3	&	2.38	&	0.19	&	12.0	\\
280794 & 2005 TV3	&	16.8	&	2.26	&	0.18	&	5.4	\\

\multicolumn{3}{l}{Hungaria candidates}   & & \\
\textbf{53424}	&	1999 SC3	&	15.2	&	1.87	&	0.06	&	23.15	\\
175122	&	2004 XM168	&	16.7	&	1.96	&	0.09	&	23.45	\\
232167	&	2002 DC10	&	16.4	&	1.93	&	0.05	&	21.06	\\

\multicolumn{3}{l}{Phocaea candidates}   & & \\
											
27842	&	1994 QJ	&	14.2	&	2.35	&	0.23	&	24.86	\\
55567	&	2002 CS6	&	13.5	&	2.32	&	0.27	&	22.95	\\
47320	&	1999 XA15	&	13.7	&	2.41	&	0.09	&	22.45	\\

\hline
\end{tabular}
\end{center}
\end{table*}

  \subsection{SpeX Observations}
    Observations were taken on the 3-meter NASA Infrared Telescope Facility at
    the Mauna Kea Observatory. We use the instrument SpeX \citep{Rayner2003},
    a near-infrared spectrograph in low resolution mode over 0.8 to 2.5
    $\mu$m.

    Objects are observed near the meridian (usually $<$ 1.3 airmass) in two
    different positions (typically denoted A and B) on a 0.8 x 15 arcsecond
    slit aligned north-south. Exposure times are typically 120 seconds, and we
    measure 8 to 12 A-B pairs for each object. Solar analog stars are
    observed at similar airmass throughout the night. We use the same set of
    solar analogs as the SMASS program \citep{Binzel2004,Binzel2006} that have been in use for
    over a decade. Uncertainties in spectral slope on the IRTF using these 
    consistent set of stars at low airmass is estimated to be around 5\% of the measured slope value.
    Observations were taken in good weather conditions and observations of 
    other objects throughout the night provide confidence that there were no 
    major systematic slope issues.

    Reduction and extraction is performed using the Image Reduction and
    Analysis Facility (IRAF) provided by the National Optical Astronomy
    Observatories (NOAO) \citep{Tody1993}. Correction in regions with strong
    telluric absorption is performed in IDL using an atmospheric transmission
    (ATRAN) model by \citet{Lord1992}. The final spectrum for each object is
    created by dividing the telluric-corrected asteroid spectrum by the
    average of the telluric-corrected solar star spectra throughout that
    night. More detailed information on the observing and reduction procedures
    can be found in \citet{Rivkin2004} and \citet{DeMeo2008}. 

  \subsection{FIRE Observations}
    Observations were taken on the 6.5-meter Magellan Telescope at Las
    Campanas Observatory. We use the instrument Folded-port InfraRed
    Echellette \citep[FIRE;][]{Simcoe2013} in high-throughput, low-resolution
    prism mode with a slit width of 0.8 arcsecond oriented toward the parallactic angle.  
    Exposures of 180 seconds were used for asteroids to avoid saturation due to thermal
    emission from the instrument and telescope at the long wavelength end (past 2.2$\mu$m).

    The readout mode sample-up-the-ramp was used for asteroid observations
    requiring exposure times in multiples of 10.7 seconds. For stars readout
    mode Fowler 2 was used. Standard stars chosen were a combination of
    well-established solar analogs used for the past decade in our IRTF
    program and newly measured G2V stars that are dimmer and better suited for
    a larger, southern hemisphere telescope.  Standard stars typically needed
    to be defocused to avoid saturation. Neon Argon lamp spectra were taken
    for wavelength calibration. Quartz lamp dome flats were taken for flat
    field corrections. 

   For FIRE data reduction, we used an IDL pipeline designed for the instrument
   based on the Spextool pipeline \citep{Cushing2004}. Among the pipeline
   settings, we use the boxcar extraction (\texttt{boxcar=1}) and local sky subtraction
   (\texttt{nolocal=1}). Tests of combinations of parameters show the best results
   with these settings. Because this slit is long (50''), for the sky correction, we use
   sky in the slit from the same exposure as the asteroid rather than
   AB pair subtractions.
    The observational circumstances are provided in Table~\ref{tab:obs}

\begin{table*}[t]
\begin{minipage}[t]{\textwidth}
\caption{Observational Circumstances and Target Information}
\label{tab:obs}
\renewcommand{\footnoterule}{}
\begin{center}
\begin{tabular}{rlcccccccccc}
\hline
  \multicolumn{2}{c}{Designation} & Phase & V & Class & Slope\footnote{Errors include only the formal error in the slope calculation.} & Albedo\footnote{Reported albedos are either from WISE \citep{Mainzer2011a,Masiero2011} or are weighted averages from WISE, IRAS \citep{Tedesco2002,Ryan2010}, and AKARI \citep{Usui2011} when multiple measurements are available.} & Est. D & H &
  SDSS & 
  \multicolumn{2}{c}{New Observations} \\
  (\#) & (Name) & (\degr) & (mag) & & (\%/1000$\AA$) & & (km) & (mag) &
  (Class)\footnote{One class is listed for each SDSS observation.} &
  (Tel.) & (Date) \\
\hline

908	&	Buda	&	13.9	&	15.2	&	D	&	6.51$\pm$0.13	&	0.114$\pm$0.025	&	29	&	10.7	&	-	&		&	2006/07/20	\\
3283	&	Skorina	&	5.1	&	15.2	&	D	&	6.39$\pm$0.03	&	0.094$\pm$0.021	&	13	&	12.6	&	D,D,X,C	&	IRTF	&	2012/07/22	\\
15112	&	Arlenewolfe	&	10.5	&	18.1	&	D	&	5.59$\pm$0.04	&	0.076$\pm$0.002	&	7	&	14.2	&	D	&	IRTF	&	2012/12/12	\\
17906	&	1999 FG32	&	19.9	&	16.6	&	D	&	5.89$\pm$0.03	&	0.072$\pm$0.009	&	10	&	13.5	&	D,D	&	IRTF	&	2012/09/16	\\
																							
&&&&&&&&&&&\\
5202	&	1983 XX	&	20.1	&	17.1	&	C/X	&	1.87$\pm$0.07	&	0.091$\pm$0.002	&	11	&	13.0	&	D	&	IRTF	&	2012/09/17	\\
14291	&	1104 T-1	&	16.4	&	17.5	&	X/D	&	2.98$\pm$0.07	&	0.102$\pm$0.018	&	7	&	13.9	&	D	&	IRTF	&	2012/09/16	\\
16025	&	1999 CA104	&	10.4	&	18	&	S	&	3.71$\pm$0.10	&	0.159$\pm$0.011	&	5	&	14.1	&	D	&	IRTF	&	2013/03/01	\\
20439	&	1999 JM28	&	1.7	&	15.9	&	S	&	1.29$\pm$0.05	&	0.331$\pm$0.015	&	4	&	13.9	&	D	&	IRTF	&	2013/01/17	\\
20452	&	1999 KG4	&	20.4	&	19.3	&	K	&	3.49$\pm$0.08	&	0.220$\pm$0.039	&	4	&	14.0	&	D	&	IRTF	&	2012/12/12	\\
33917	&	2000 LK19	&	3.6	&	17.9	&	C/X	&	1.18$\pm$0.06	&	0.114$\pm$0.015	&	4	&	14.9	&	D	&	IRTF	&	2013/03/01	\\
53424	&	1999 SC3	&	24.1	&	17.9	&	S/L	&	2.62$\pm$0.08	&	-	&	-	&	15.2	&	D	&	IRTF	&	2012/12/13	\\
125102	&	2001 UH35	&	23.3	&	19.1	&	X/D	&	2.87$\pm$0.15	&	-	&	-	&	15.3	&	D,D,K	&	IRTF	&	2013/03/01	\\
224306	&	2005 UV8	&	8.6	&	18.3	&	S	&	2.18$\pm$0.07	&	-	&	-	&	16.1	&	D,D	&	Magellan	&	2012/08/08	\\
247264	&	2001 SW8	&	8.1	&	18.1	&	C/X	&	2.17$\pm$0.08	&	0.065$\pm$0.024	&	-	&	16.3	&	D,D	&	IRTF, Magellan	&	2012/09/16	\\
\hline
\end{tabular}
\end{center}
\end{minipage}
\end{table*}

\section{Results}

  We confirmed 3 D-type inner belt asteroids from SDSS follow-up
  observations. Of SDSS
  candidates that had multiple observations 2 out of 5 were found to be
  D (40\% confirmation success).
  For singly observed SDSS candidates 1/7 was confirmed (15\%).
  We note that
  asteroids (14291) and (125102) had spectral slopes that put them on the
  border of the X and D classes (if they were to be considered as D, the
  fraction confirmed would raise to 60\% and 28\% respectively). 
  There is ambiguity
  between classes particularly for slopes near the boundaries of two classes.
  While formal classifications boundaries have been defined \citep[e.g.,][]{Tholen1984,Bus2002b,DeMeo2009taxo}
  these strict boundaries are in some cases artificial. Additionally,
  observational effects, such as airmass differences between object and solar analog, phase angle of the observations,
  and loss of light in the slit, cause spectral slope uncertainties on the order of 5\% that could cause
  an object on the border to cross over to another class.
  For this reason we choose to include only the most distinctly D-type spectra
  where there is no ambiguity and thus do not consider
  the two border cases as D-types in order to be most conservative in assigning an asteroid
  to the D class. We thus provide a lower limit to the number of D-types in the inner belt.

  We find all the confirmed D-types have H magnitudes less than 15, while
  all 4 candidates with H greater than 15 were false positives. Because of
  the small size of our sample, we do not know if this indicates fewer
  D-types at smaller sizes or is merely a reflection of a higher false
  positive rate for presumably lower SNR for data of the smaller, dimmer
  objects. 
  
\begin{figure}
  \centering
  \includegraphics[width=\textwidth]{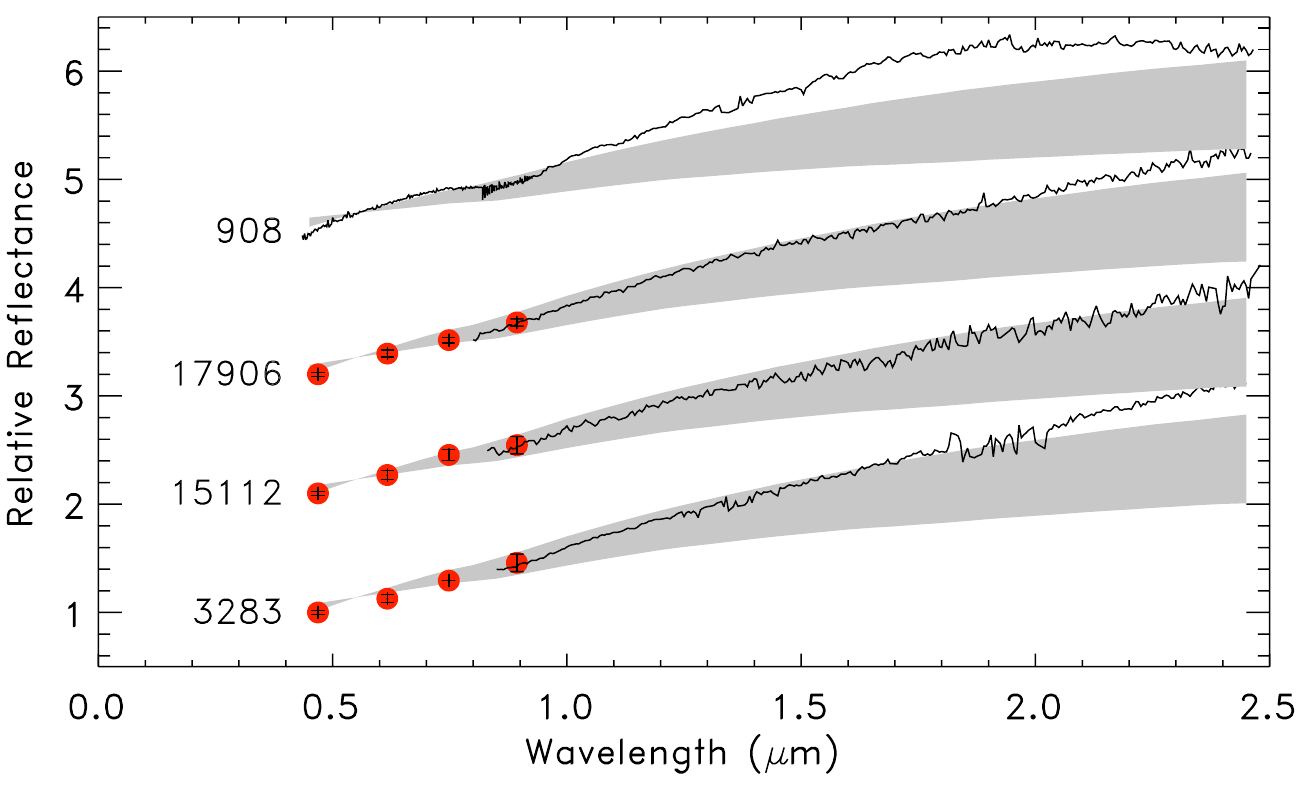}
  \caption[Spectra of confirmed D-types]{%
  Plot of confirmed D-types based on near-infrared spectral measurements from
  this work for three objects plus asteroid (908) from
  \citep{DeMeo2009taxo}. SDSS colors are plotted as red dots with black error bars (smaller than the 
  size of the dot). The spectra are plotted in black. The gray region bounds plus and minus one sigma
  from the mean of the D-type class. The three SDSS D-types have nearly identical
  spectra.
    } 
  \label{fig: specd}
\end{figure}

  Fig.~\ref{fig: specd} plots the spectra of confirmed inner belt D-type
  asteroids. The three SDSS candidates look strikingly similar. The spectra of
  (3282) Spencer Jones,
  (15112) 2000 EE$_{17}$, and
  (17906) 1999 FG$_{32}$
  have slopes (between 0.85 and 2.45 $\mu$m) of
  6.4, 5.6, and 5.9 \%/1000$\AA$, respectively which is very comfortably within the
  D-type slope range of $\gtrsim$2.5\%/1000$\AA$ \citep{DeMeo2009taxo}.

\begin{figure}
  \centering
  \includegraphics[width=\textwidth]{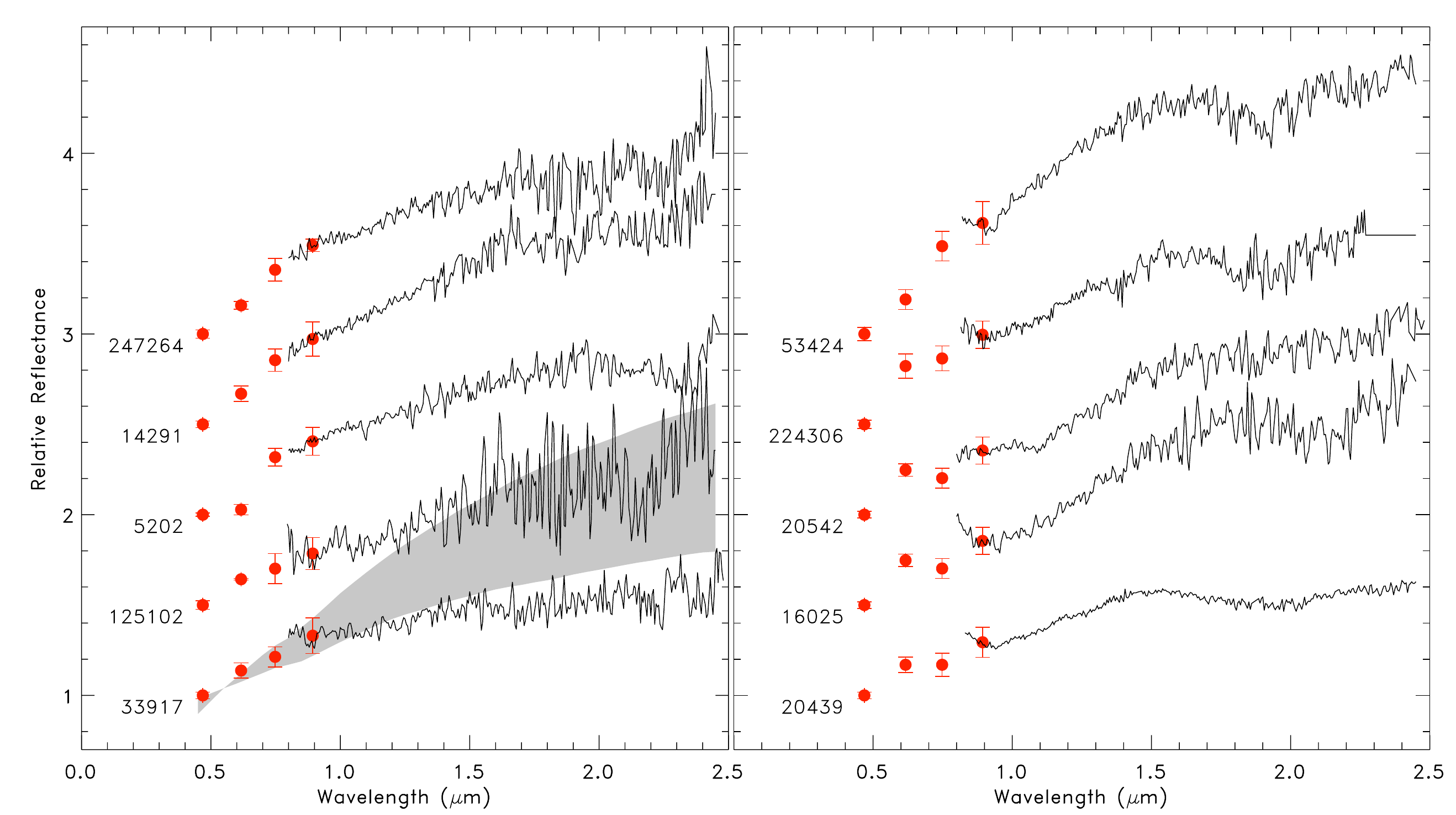}
  \caption[Spectra of rejected D-types candidates]{%
  Plot of SDSS candidates that are not D-types based on near-infrared follow
  up observations. SDSS colors are plotted as red dots with red error bars. 
  The spectra are plotted in black. The gray region plotted with the 
  first spectrum shows the boundaries plus and minus one sigma
  from the mean of the D-type class illustrating the difference between D-types and these spectra.
  On the left are featureless or subtly featured spectra with
  slopes that place them in the C or X complex. On the right are spectra with
  1 and 2 $\mu$m features that place them in the S, K, or L classes. 
    } 
  \label{fig: false}
\end{figure}

  The SDSS D-type candidates that were false positives are plotted in
  Fig.~\ref{fig: false}~and exhibit a wide range of spectral types. For many of 
  these objects the false positives could be due to real slope difference between
  the visible and near-infrared. The asteroids in the right panel of Fig.~\ref{fig: false}~have distinct 
  one- and two- micron absorption features indicating the presence of the minerals olivine and pyroxene,
  placing these spectra in the S, K, and L classes. We suspect
  that for these poorer matches, the false positive is due to the noise level of the
  SDSS measurement for that particular object and not due to a serious mismatch in
  the visible and near-infrared spectral characteristics.

Beyond the candidates in this survey, we find one asteroid with spectral measurements in both the
      visible and near-infrared range that place it in the D class:
      (908) Buda \citep{DeMeo2009taxo}.
        Asteroid (908) Buda, taken as part of
  the SMASS survey \citep{Bus2002a,DeMeo2009taxo}, was classified as an L-type
  by the visible wavelength survey, whereas the near-infrared data are
  inconsistent with an L-type, putting it in the D class.
  The orbital parameters of this object had not
  been reexamined after the classification change until now. 
      This object has a moderately low to medium albedo of 0.087\,$\pm$0.007
      \citep{Masiero2011} from WISE and 0.16\,$\pm$0.02 \citep{Tedesco2002} or 0.14\,$\pm$0.01 \citep{Ryan2010} from IRAS
      and a large spectral slope.      
     The spectrum and two interesting features: a subtle absorption band centered
      near 0.9 microns and a distinct flattening of the spectrum to a more
      neutral slope past 2 microns. 
      While the spectral slope clearly places (908) Buda in the D-class its
      moderate albedo would place it in the M class in the Tholen system.
      Similar subtle features on other asteroids are attributed to low iron, low calcium orthopyroxene \citep{OckertBell2008,Fornasier2010},
      although these objects fall in the Tholen M class or Bus-DeMeo Xk class. 
      The visible wavelength survey by 
      \citet{Lazzaro2004} detected 5 inner belt D-types, asteroids (565), (732), (1689), (2105) and (4103),
      the orbits of which had not been examined until now. Most of their albedos, however, are also significantly larger 
      than typical for D-types: 0.11$\pm$0.02, 0.14$\pm$0.02, 0.18$\pm$0.05, 0.16$\pm$0.03, and 0.25$\pm$0.06 \citep{Masiero2011}.

\section{Comparison with NEOs}

In the SMASS \citep{Binzel2004,Binzel2006} dataset of near-Earth objects (NEOs),
there are 12 known D-types. While it had traditionally been assumed
that these D-type NEOs come from the Jupiter Family Comet region or
possibly the outer belt, source region probabilities derived from the model
by \citet{Bottke2002} suggest that a significant number of these D-types (7/12)
have orbits that make them highly likely to have originated in the middle or inner belt.
All 7 of these NEOs have semi-major axes less then 2.5 AU, while those likely to have
originated in the outer belt or from the Jupiter Family Comets have semi-major axes greater
than 2.5 AU.
Table~\ref{tab:neos}
provides a probability estimate for an NEO's origin from multiple sources: a Jupiter Family Comet (JFC),
outer asteroid belt (OB), 3:1 mean-motion resonance (3:1), Mars Crosser (MC), and the $\nu_6$ resonance (N6).
While the relative abundance of D-types in the inner belt is much lower than for the outer regions, the 
3:1 and $\nu_6$ resonance delivery mechanisms are significantly more efficient, meaning it could be possible
that some of the NEO D-types actually originate from the inner belt.

\begin{table*}[t]
\begin{minipage}[t]{\textwidth}
\caption{Source Region Probabilities for D-type NEOs}
\label{tab:neos}
\renewcommand{\footnoterule}{}
\begin{center}
\begin{tabular}{lcccccccccc} \hline

\hline
Asteroid	&	Orbit\footnote{NEO orbits are labeled as Amor (AMO), Apollo (APO) and Aten (ATE)}	&	a (AU)	&	e	&	i (deg)	&	T$_J$	&	 P$_{JFC}$	&	P$_{OB}$	&	P$_{3:1}$	&	P$_{MC}$	&	P$_{\nu_6}$	\\
\hline
52762	&	APO	&	2.419	&	0.651	&	33.9	&	3.01	&	0.000	&	0.034	&	0.734	&	0.159	&	0.072	\\
162998	&	AMO	&	1.926	&	0.474	&	1.7	&	3.77	&	0.000	&	0.095	&	0.053	&	0.202	&	0.650	\\
170891	&	AMO	&	1.983	&	0.405	&	8.1	&	3.74	&	0.000	&	0.000	&	0.032	&	0.299	&	0.670	\\
2003 UC20	&	ATE	&	0.781	&	0.337	&	3.8	&	7.39	&	0.000	&	0.000	&	0.145	&	0.332	&	0.523	\\
2005 DD	&	APO	&	1.933	&	0.568	&	7.3	&	3.69	&	0.000	&	0.000	&	0.073	&	0.324	&	0.603	\\
2006 MJ10	&	APO	&	1.876	&	0.586	&	39.3	&	3.53	&	0.000	&	0.000	&	0.331	&	0.290	&	0.379	\\
2013 AH11	&	AMO	&	2.274	&	0.521	&	28.3	&	3.28	&	0.000	&	0.000	&	0.856	&	0.055	&	0.089	\\
	&		&		&		&		&		&		&		&		&		&		\\
3552	&	AMO	&	4.222	&	0.713	&	31.0	&	2.31	&	1.000	&	0.000	&	0.000	&	0.000	&	0.000	\\
17274	&	AMO	&	2.725	&	0.558	&	5.6	&	3.10	&	0.028	&	0.750	&	0.071	&	0.144	&	0.008	\\
326732	&	AMO	&	2.705	&	0.575	&	6.3	&	3.10	&	0.028	&	0.750	&	0.071	&	0.144	&	0.008	\\
2000 PG3	&	APO	&	2.824	&	0.856	&	22.0	&	2.55	&	0.929	&	0.025	&	0.037	&	0.002	&	0.007	\\
2011 BE38	&	APO	&	2.620	&	0.719	&	7.9	&	2.96	&	0.438	&	0.346	&	0.128	&	0.064	&	0.025	\\
\hline

\end{tabular}
\end{center}
\end{minipage}
\end{table*}

\section{Bias-corrected abundance and distribution}
  We seek to use our observed sample to determine the total number of D-types
  expected to exist in the inner main belt. Here we exclude the Hungarias and Phocaeas 
  because we do not have a large enough sample to determine the follow-up
  success rate in those regions. The SDSS survey
  is efficient to H magnitudes of about 17 in the inner belt \citep{Ivezic2001,DeMeo2013}.
  We determine the
  expected total number of D-types in each H magnitude bin for the
  SDSS sample (N$_{\rm SDSS, D}$) by first multiplying the 
  number of SDSS D-type candidates by the success rate (1/6 or 2/5) based on the number
  of singly or multiply observed objects in each bin. Note we use 1/6 because we exclude
  the one Hungaria asteroid that was observed. We then determine the number
  of inner belt objects (N$_{\rm SDSS, total}$) from the entire SDSS sample
  over that size range in the inner belt
  and then calculate what fraction the D-types represent (D fraction).
  We can then apply 
  that fraction to the total number of objects that exist
  (N$_{\rm MPC}$) to find the total number of inner belt D-types (N$_{\rm MPC, D}$). The AstOrb database
  hosted at the Minor Planet Center (MPC) is essentially complete in the inner
  belt to H=16 \citep{DeMeo2013}.
  The MPC is 71\% complete in the inner belt between 16
  $<$ H $<$ 17, so we apply a correction factor to account for the expected undiscovered objects. 
  
  We place upper limits equal
  to the mean of the Poisson distribution for which there is a 90\% chance of
  observing n+1 or more candidates where n is the number of expected objects
  in a given H bin of our SDSS sample (for H bins between H of 12 and 16 n is 1,2,4,4,3,
  respectively) as was done in similar work for V-types by
  \citet{Moskovitz2008}. The lower limit is equal to the mean of the Poisson
  distribution for which there is a 90\% chance of observing n-1 or fewer
  candidates. Table~\ref{tab:debias} provides the total number of D-types in the
  inner belt for each H magnitude bin. 
  
  Even though the SDSS survey is efficient to an H magnitude of 17, it is severely biased
  against asteroids with H$<$12 because they saturated the detector during observations. The spectral
  surveys make up for this bias because they are nearly 100\% complete down to H=12 in the inner belt \citep{DeMeo2013}.
  Five D-types were identified in the inner belt from \citeauthor{Lazzaro2004} based on visible wavelengths.
  Among the sample of asteroids observed at visible plus near-infrared wavelengths, all near-ir D-types have
  high slopes in the visible, but the reverse is not always true for D-types with only visible measurements \citep{DeMeo2009taxo}. 
  Additionally, the albedos for these targets are significantly higher than for typical D-types.
  For these reasons we do not include them in our calculations for this work, but if near-infrared observations confirms
  their very red slopes continue to the infrared, they should be included more quantitatively among the sample.
  
\begin{table*}[t]
\begin{minipage}[t]{\textwidth}
\caption{Observational Circumstances and Target Information}
\label{tab:debias}
\renewcommand{\footnoterule}{}
\begin{center}
\begin{tabular}{lllllll} \hline

\bf H	&
\bf N$_{\rm SDSS, D}$\footnote{The number of inner belt
  D-types expected statistically in the SDSS sample calculated as the
  detection rate times the number of D-type candidates from the SDSS
  sample.}	&
\bf N$_{\rm SDSS, total}$\footnote{The total number
  of objects in the inner belt observed in our SDSS sample.} 	&
\bf D fraction\footnote{The fraction of D-types at each H magnitude
  range calculated as N$_{SDSS, D}$/N$_{\rm SDSS, total}$}	&	\bf
  N$_{\rm MPC}$\footnote{The total number of inner belt asteroids at
  each H magnitude range from the MPC with a correction for
  incompleteness in the last bin.}	&	
\bf N$_{\rm MPC, D}$\footnote{The total number of inner belt D-types expected
  statistically. The error bars are based on the Poisson upper and
  lower limits as described in the text.}\\ 
\hline
12-13	&	0.4	&	161	&	0.00248	&	712	&	1.8$^{+8.7}_{-1.6}$	\\
13-14	&	2.3	&	847	&	0.00248	&	3185	&	8.6$^{+31.6}_{-6.5}$	\\
14-15	&	4.1	&	2209	&	0.00167	&	10236	&	19.0$^{+57.0}_{-10.9}$	\\
15-16	&	4.5	&	3860	&	0.00106	&	24928	&	29.1$^{+87.2}_{-16.7}$	\\
16-17	&	3.5	&	3635	&	0.00085	&	69368	&	66.8$^{+215.7}_{-42.1}$	\\
Total:	&		&	10712	&		&	108429	&	125.3	\\

\hline

\end{tabular}
\end{center}
\end{minipage}
\end{table*}

\section{Discussion}

\begin{figure}
  \centering
  \includegraphics[width=\textwidth]{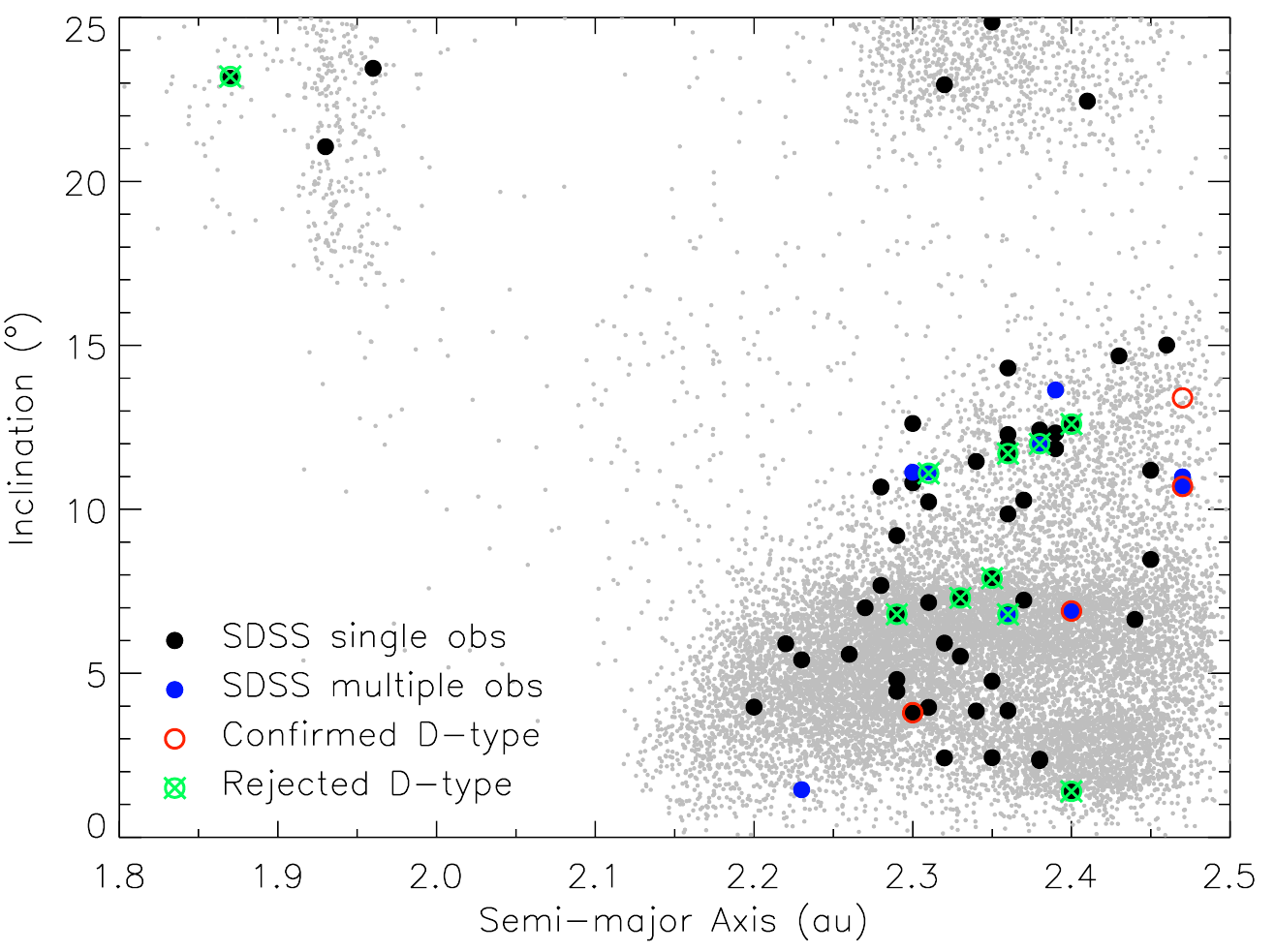}
  \caption[Orbital distribution of D-types in the inner belt]{%
  Orbital distribution of D-types in the inner belt. This plot shows the location
  of the candidate D-types located closer than 2.5AU. The bottom right region
  is the inner main belt. The top left is the Hungarias and the top right is the
  Phocaeas. A sample of MBAs are plotted in gray to illustrate the structure
  of the region. SDSS candidates are plotted as black or blue dots for
  objects with single or multiple observations respectively. Red circles
  indicate D-types that have been confirmed with near-infrared data.
  Green X marks objects with follow up observations that do not classify as D-types.
    } 
  \label{fig: orbd}
\end{figure}

  Fig.~\ref{fig: orbd} plots the orbital elements of the SDSS D-type
  candidates and the objects observed in this work. The candidates and
  confirmed D-types display a wide range of orbits throughout the inner belt
  suggesting these D-types do not all originate from a single source or
  location such as being disrupted fragments of an originally larger parent
  body. We note that among the larger, visible-wavelength D-types from
  \citet{Lazzaro2004} (not plotted), two are located among the Phocaeas and the other three
  are at the outer edge of the inner belt around 2.45AU.

  Albedos listed in Table~\ref{tab:obs} are a weighted average from WISE \citep{Mainzer2011a}, 
  IRAS \citep{Tedesco2002,Ryan2010}, and AKARI \citep{Usui2011}
  when available or from WISE when that is the only survey with data. The
  median albedo of our inner belt D-types is 0.09, with values ranging from 0.07 to
  0.11. This is consistent with albedos found for Bus D-types although slightly higher
  than the albedos for most Tholen and Bus-DeMeo D-types \citep{Mainzer2011}.

  We estimate there are on the order of 100 inner belt D-type asteroids 
  with H magnitudes between 12 and 17 which corresponds to diameters between 2.5 and 20km.
  While we do
  not follow up D-type SDSS candidates in the middle belt, we find 184 
  candidates. Assuming a similar confirmation rate we expect there to be on
  the order of 200-250 middle belt D-types over the same H magnitude
  range. Assuming an average albedo and density we can calculate the total
  volume and estimated mass of D-type material in the inner belt. For a lower limit we use an
  albedo of  0.09, which is the median of the D-types observed in this work and a
  density of 1.0 g/cm$^3$. This density agrees with the single D-type density
  measurement available which is 1.0$\pm$0.02 for asteroid (624) Hektor \citep{Marchis2013} 
  and is also consistent with densities of comets and
  transneptunian objects \citep{Carry2012}.  We find the lower limit to the total volume and mass
  to be 2.6x10$^{13}$m$^3$ and 2.6x10$^{16}$kg which represents $\sim$0.01\% of the mass of the
  inner belt. We calculate an upper limit for the mass assuming a lower albedo (0.04) and 
  a density of 1.8g/cm$^3$ which is the at the upper end of the range for C-types \citep{Carry2012}.
  In this case the mass increases by nearly an order of magnitude. If (908) Buda and the 5 D-types from \citet{Lazzaro2004}
  are included, which range in diameter from 15 to 30km, that would increase the mass by
  another factor of a few assuming the 0.09 albedo and 1.0 g/cm$^3$ density case. 
  By mass, D-types represent roughly 1, 2, 15, 67\% of the outer belt,
  Cybeles, Hildas and Trojans \citep{DeMeo2013}. Inner belt D-types represent,
  as expected, a very small fraction of the inner belt as well as a small
  fraction of the total D-type population. 

  As mentioned in the Results Section, we did not confirm any D-types with H$>$15.
  If, in fact, there are fewer D-types at the smallest sizes we sample, that would
  change the number of D-types greater than 2 km from $\sim$100 objects 
  to $\sim$25 (see Table~\ref{tab:debias}), although the total mass would not change substantially.

  The currently favored theory of solar system evolution includes 
  periods of planetary migration that displace large numbers of small bodies
  \citep[e.g.][]{Gomes2005,Morbidelli2005,Tsiganis2005,Walsh2011}.  
  \citet{Levison2009} produced simulations showing that P- and D-type
  asteroids originating in the outer solar system could have been implanted
  into the Trojans and could have reached as far as the outer belt. At that
  time no P- or D-types had been observed in the inner or middle parts of the
  belt.  P-type is a taxonomic class in the Tholen taxonomy \citep{Tholen1984} with relatively 
  featureless and moderately red spectra and low albedos, that falls within the X-complex in the Bus-DeMeo taxonomy \citep{DeMeo2009taxo}. 
  Because C and P type asteroids are less easily
  distinguished especially in the near-infrared range we prefer to focus on
  the more spectrally distinct D-types. We find 3 D-types in the inner belt,
  the largest of which has a diameter of about 10\,km and 1 anomalous D-type (908) Buda 
  with a diameter of about 30km. The work by
  \citet{Levison2009} only includes objects with diameters greater than 40\,km.
  Because of the size mismatch between our observations and the simulations,
  perhaps some of the more numerous smaller bodies
  reached farther distances inward or these smaller bodies have
  managed to move from other regions of the belt later on in solar system history.
  Dynamical models have yet to show how D-types could penetrate into the inner reaches 
  of the Main Belt from the Kuiper Belt under current scenarios of planet formation.
  
If these objects were originally implanted in the outer belt, or even in the middle belt, the D-type asteroids 
would still need to cross major resonances, particularly the 3:1. The major resonances have a strong 
eccentricity pumping effect acting on short (typically $<$1My) timescales, pushing the objects out of the 
main belt and into planet crossing space \citep{Gladman1997}. However, there are at least two plausible 
ways a body could cross. First, a single energetic event such as the break up of a parent body into a family
 near the 3:1 that could provide enough energy to quickly cross. The varied orbits of the inner belt D-types 
 and lack of evidence for a remnant family on the other side of the 3:1 makes this scenario unlikely. 

Alternatively, the Yarkovsky force causes objects with diameters on the order of tens of km or smaller to drift 
slowly in semi-major axis space. \citet{Nesvorny2008} investigate the dynamical spread of the Vesta family 
members over 2Gy and found that none of their 132 test objects that entered the 3:1 resonance succeeded 
to cross it. In fact, all were removed from the main belt by eccentricity pumping putting them on planet-crossing 
orbits. \citet{Roig2008} also examined the possibility of Vesta family members crossing the 3:1 resonance and 
found that the $\sim$5km Vesta-like asteroid (21238) located in the middle belt has only a $\sim$1\% probability 
of having originated from the Vesta family. They find, however, that objects smaller than 5km have a much higher 
probability of resonance crossing. While many of the inner belt D-types are larger than 5km, they also have much
lower albedos than the Vesta-like objects studied in previous work and are thus generally more affected by Yarkvosky 
drift \citep{Vokrouhlicky2001}. Nevertheless, the Yarkovsky effect is dependent on many factors including thermal inertia, shape, density, 
rotation state and other physical properties. Further work is needed to investigate the efficiency of objects 
crossing resonances particularly as a function of size and albedo.

In another scenario, these D-types could have been transported during another migration period.
Most of the P- and D-types were thought to have been transported during a late-stage migration caused by 
interactions between the giant planets and the Kuiper Belt that destabilized the region and sent bodies inward
to the inner solar system \citep{Levison2011}. However, in a hypothesized early stage migration, material near and between the
giant planets is moved inward when Jupiter migrates inward to about 1.5 AU and then back outward \citep{Walsh2011}.
We would expect most of this material to have been C-type, but it is plausible that there was compositional
variation among that population that would cause some these bodies to look more like a D-type.

  The inner belt D-types have albedos at or above the upper limit of the typical range for D-type 
  asteroids in the outer belt to the Trojans  \citep[0.03-0.07, ][]{Fernandez2003,Mainzer2011,DeMeo2013}.
  The objects in our sample are much smaller than the typical sizes of D-types measured further 
  out in the solar system. Smaller sizes typically indicate
  a younger surface age, so it is possible the difference is due to weathering effects.
  Because of the lack of distinguishing features for D-types, it is also possible that the
  objects we find in the inner belt are 
  compositionally distinct from other D-types. 
  
   Inner belt D-types have a size-frequency distribution drastically different from other regions. All confirmed and
   candidate D-types in the inner belt
  have D$\lesssim$30km which have dynamical lifetimes expected to be shorter than
  the age of the solar system. There are none in the medium (30-100km) or large ($>$100km) size range. 
  This means they are most likely collisional fragments of a larger body,
  however there are no candidate D-type parents in the inner belt. If these D-types originated from another
  D-type, then the parents are either completely destroyed or were never in the inner belt at all. The presence 
  of these bodies also raise the question as to whether they could originate from a larger parent of a different 
  taxonomic type.

\section{Conclusion}

  We identified inner belt D-type candidates from the Sloan Digital Sky Survey
  Moving Object Catalog and confirmed 3 D-types from new near-infrared
  spectral observations. We estimate there are
   $\sim$100 and $\sim$250 D-types with diameters between 2.5
    and 20km in the inner and middle belt, respectively. The average albedo of
    the inner belt D-types (0.09) is slightly higher than for typical D-types.
  D-types are thought to originate
  from the outer solar system, however, models by \citet{Levison2009} show
  that they are not expected to reach as far as the inner belt. There are many possible scenarios
  that could explain the presence of these inner belt D-types: 
  i) They were scattered farther than expected during the late-stage migration modeled by \citet{Levison2009}.
  ii) They arrived through another mechanism such as an earlier migration, other planetary scattering, or
  Yarkovsky drift across the resonances.
  iii) They are compositionally distinct from other D-types and thus do not require an implantation mechanism.

  The sample size in this work is small, but has
  uncovered an important new population in the inner belt. Future work discovering inner belt D-types
  with measured visible and near-infrared spectra and albedos will help us understand the frequency and origin
  of these bodies.


\section*{Acknowledgments}
  We thank Bill Bottke for discussions and providing source region probabilities for the NEOs.
  Observations reported here were obtained at the NASA Infrared Telescope
  Facility, which is operated by the University of Hawaii under Cooperative
  Agreement NCC 5-538 with the National Aeronautics and Space Administration,
  Science Mission Directorate, Planetary Astronomy Program.  
  This paper includes data gathered with the 6.5 meter Magellan Telescopes
  located at Las Campanas Observatory, Chile. 
  We acknowledge support from the Faculty of the European Space Astronomy 
  Centre (ESAC) for FD's visit. DP is grateful to the AXA research fund.
  This material is based upon work supported by the National Aeronautics and
  Space Administration under Grant No. NNX12AL26G issued through the Planetary
  Astronomy Program and by the National Science Foundation under Grant
  0907766. Any opinions, findings, and conclusions or recommendations
  expressed in this article are those of the authors and do not necessarily
  reflect the views of the National Aeronautics and Space Administration or
  the National Science Foundation.

\clearpage
\bibliographystyle{elsarticle-harv.bst}

\end{document}